\documentclass[12pt]{article}
\usepackage{graphicx}

\newcommand{\initiate}{\setcounter{equation}{0}}
\newcommand{\ls}[1]{#1\hspace{-.21cm}/}
\usepackage{amsmath}
\usepackage{amsbsy}
\usepackage{amsfonts}
\usepackage{amsthm}
\begin{document}

\title{Q-stars in $2+1$ dimensions}

\author{Athanasios Prikas}

\date{}

\maketitle

Physics Department, National Technical University, Zografou
Campus, 157 80 Athens, Greece.\footnote{e-mail:
aprikas@central.ntua.gr}

\begin{abstract} We study q-stars with one or two scalar fields,
non-abelian, and fermion-scalar q-stars in $2+1$ dimensions in an
anti de Sitter or flat spacetime. We fully investigate their
properties, such as mass, particle number, radius, numerically,
and focus on the matter of their stability against decay to free
particles and gravitational collapse. We also provide analytical
solutions in the case of flat spacetime and other special cases.
\end{abstract}

PACS number(s): 11.27.+d, 04.40.-b

\newpage

\section{Introduction}

The investigation of scalar field configurations coupled to
gravity started with the work by Kaup, \cite{kaup}, and Ruffini
and Bonazzola, \cite{ruf}. They regarded a complex scalar field
with no self-interactions. Quartic self interactions were taken
into account in other works \cite{quartic1,quartic2}. Scalar
fields with a local $U(1)$ symmetry were also investigated
\cite{charged}. The results were generalized in a series of papers
in other gravity theories \cite{tor1,tor2,tor3,tor4}.

Soliton stars appeared in the literature as stable field
configurations consisted of abelian or non-abelian scalar fields,
containing sometimes fermions or gauge bosons coupled to gravity
\cite{fr1,fr2,fr3,lyn1,lyn2,lyn3,lyn4,brito}. These
configurations are stable even in the absence of gravity. A
certain class of soliton stars are very large field configurations
with radius of the order of lightyears \cite{fr1,fr2,fr3}. They
consist of a very large interior with very low energy density of
``kinetic" type resulting from the time variation of the scalar
field. On the other hand the potential energy within the soliton
is considered to be negligible. Also, the metric and matter
fields vary very slowly in the soliton interior. A considerable
amount of energy, comparable to that stored in the huge interior,
is contained in the thin surface. Within the surface both the
potential and the energy resulting from the radial variation of
the matter field contribute to the total surface energy, but the
contribution of the time variation of the field is very small, as
it is supposed to rotate in its internal $U(1)$ space extremely
slowly.

Another class of non-topological soliton stars, the q-stars,
appeared as relativistic generalizations of a certain family of
non-topological solitons, namely q-balls. Q-balls have a certain
role in particle theory and especially in the baryogenesis
through the flat MSSM directions \cite{dine}. They appear in
Lagrangians, \cite{col1,col2,tracas,kusenco}, with a global $U(1)$
or $SU(3)$, $SO(3)$ symmetry, or a local $U(1)$ \cite{local},
when the scalar field takes the special value which minimizes the
$U/{\phi}^2$ quantity.

The properties of the q-stars have been studied thoroughly in
\cite{lyn1,lyn2,lyn3,lyn4,lyn5,pr}. There are q-stars with one or
two scalar fields, \cite{lyn1,pr}, q-stars with non-abelian
fields, \cite{lyn5}, and q-stars with a fermion and a scalar
field, \cite{lyn2,lyn3,lyn4}. These objects contain a large
interior within which both metric and matter fields vary
smoothly. The energy within the interior is the sum of the
potential energy and of the ``kinetic" energy resulting from the
time variation of the matter field. In the case of q-balls and,
consequently, the case q-stars the potential is always positive.
We will name $U$, $W$ and $V$ the potential energy, the energy
resulting from the time variation and the energy resulting from
the space variation respectively. There is a crucial relation
that roughly describes the q-solitons and differentiates them
from other classes of non topological solitons, namely:
\begin{equation}\label{0.1}
\omega\sim\phi\sim m\ ,
\end{equation}
where $\omega$ is the frequency with which the soliton rotates in
its internal $U(1)$ space, or, with proper generalizations that
we describe, in more complicated spaces, $\phi$ is a typical
value of \textit{the} or \textit{one} of the scalar field(s)
within the soliton and $m$ is the mass of the free particles.
This relation means that $W$ is of the same order of magnitude
with $U$, in contrast with other non-topological solitons for
which the $U$ is negligible in the interior and $W$ negligible
within the surface. The common feature between q-stars and other
soliton stars is the smooth variation of the matter field in the
interior and the rapid change, within the surface, from a
${\phi}_0$ value to zero. In non-topological soliton stars, the
energy density within the surface is huge compared to the energy
density in the interior, and so, although the surface is very
thin, the energy, contained within it, is comparable to the energy
stored in the interior. This does not hold true for the q-stars.
They have an approximately constant energy density in the
interior, the sum of $W$ and $U$. At the surface, these
quantities retain the same order of magnitude and $V$ in now
added with the same order of magnitude as well. Because the
energy density is everywhere of the same order of magnitude,
either in the interior, or in the surface, and because the
surface is of width of $m^{-1}$, the total energy of the thin
surface is negligible. It has been proved that q-stars are
smaller than non-topological soliton stars but larger than boson
stars with no soliton features in the absence of gravity.

Gravity in $2+1$, \cite{des1,des2}, dimensions coupled to scalar
fields is studied for many reasons. One of them is the
theoretical interest for a theory qualitatively different from
the corresponding theory in $3+1$ dimensions. The solutions in a
$3-$dimensional theory of gravity are much more simpler because
there is no $A/r^2-1$ term in the Einstein tensor, where $A$ is a
metric field as we will see. This can lead even to analytical
solutions for special field configurations composed of scalar
fields (or fermion-scalar) coupled to $3-$dimensional gravity
\cite{shi1,shi2,shi3}.

The aim of the present article is to study the formation of
q-stars in $2+1$ dimensions in anti anti de Sitter or flat
spacetime, as a limiting case. We want to investigate the matter
of their stability with respect to gravitational collapse and to
fission into free particles and to study the influence of the
global spacetime curvature in the features of the q-star. We study
four different kinds of q-stars, namely, with one and two scalar
fields, non-abelian q-stars and fermion-scalar q-stars. We
thoroughly investigate their properties, such as soliton radius,
mass, particle number, as functions of either the cosmological
constant, which is an absolutely independent parameter, or as
functions of their internal frequency, a quantity with crucial
role in the theory of non-topological solitons. We also give
analytical solutions for the important case that the cosmological
constant tends to zero and also for other cases, which show a
very close similarity in their behavior, when compared to
solutions obtained with the usual numerical methods.

\initiate
\section{A q-star with one scalar field}

We consider a static, spherically symmetric metric:
\begin{equation}\label{1.1}
ds^2=-e^{\nu}dt^2+e^{\lambda}d{\rho}^2+{\rho}^2d{\alpha}^2\ ,
\end{equation}
with $g_{tt}=-e^{\nu}$. We regard a scalar field with the
minimum-energy time dependence:
\begin{equation}\label{1.2}
\phi(\vec{\rho},t)=\sigma(\rho)e^{-\imath\omega t}\ .
\end{equation}
The action in natural units for a scalar field coupled to gravity
in 2+1 dimensions is:
\begin{equation}\label{1.3}
S=\int d^3x\sqrt{-g}\left[\frac{R-2\Lambda}{16\pi G}+g^{\mu\nu}
{({\partial}_{\mu}\phi)}^{\ast}({\partial}_{\nu}\phi)-U\right]\ ,
\end{equation}
where $\Lambda$ stands for the cosmological constant, regarded
here to be negative, or zero, as a limiting case. The
energy-momentum tensor is:
\begin{equation}\label{1.4}
T_{\mu\nu}={({\partial}_{\mu}\phi)}^{\ast}({\partial}_{\nu}\phi)+
({\partial}_{\mu}\phi){({\partial}_{\nu}\phi)}^{\ast}
-g_{\mu\nu}[g^{\alpha\beta}{({\partial}_{\alpha}\phi)}^{\ast}({\partial}_{\beta}\phi)]
-g_{\mu\nu}U\ .
\end{equation}
The theory need not be fundamental and, thus, renormalizable but
effective. The Euler-Lagrange equation for the matter field is:
\begin{equation}\label{1.5}
\left[1/\sqrt{|g|}{\partial}_{\mu}(\sqrt{|g|}g^{\mu\nu}{\partial}_{\nu})-
\frac{dU}{d{|\phi|}^2}\right]\phi=0\ ,
\end{equation}
taking now the form:
\begin{equation}\label{1.6}
{\sigma}''+[1/\rho+(1/2)({\nu}'-{\lambda}')]{\sigma}'+e^{\lambda}
{\omega}^2e^{-\nu} \sigma
-e^{\lambda}\frac{dU}{d{\sigma}^2}\sigma=0\ .
\end{equation}
The Einstein equations are:
\begin{equation}\label{1.7}
G_{\ \nu}^{\mu}\equiv R_{\ \nu}^{\mu}-\frac{1}{2}{\delta}_{\
\nu}^{\mu}R=8\pi G T_{\ \nu}^{\mu}-\Lambda \delta^{\mu}_{\ \nu}\ .
\end{equation}
With the above assumptions the Einstein equations take the form:
\begin{equation}\label{1.8}
-e^{-\lambda}\frac{\lambda'}{2\rho}=8\pi G(-W-V-U)-\Lambda\ ,
\end{equation}
\begin{equation}\label{1.9}
e^{-\lambda}\frac{\nu'}{2\rho}=8\pi G(W+V-U)-\Lambda\ ,
\end{equation}
where
\begin{equation}\label{1.10}
\begin{split}
W\equiv e^{-\nu}{\left(\frac{\partial\phi}{\partial
t}\right)}^{\ast}\left(\frac{\partial\phi}{\partial t}\right)=
e^{-\nu}{\omega}^2{\sigma}^2, \\ V\equiv
e^{-\lambda}{\left(\frac{\partial\phi}{\partial\rho}\right)}^{\ast}
\left(\frac{\partial\phi}{\partial\rho}\right)=
e^{-\lambda}{\sigma'}^2\ .
\end{split}
\end{equation}

There is a Noether current:
\begin{equation}\label{1.11}
j^{\mu}=\sqrt{-g}g^{\mu\nu}\imath({\phi}^{\ast}{\partial}_{\nu}\phi
-\phi{\partial}_{\nu}{\phi}^{\ast})\ .
\end{equation}
The current is conserved according to the equation:
\begin{equation}\label{1.12}
j^{\mu}_{\ ;\mu}=0\ .
\end{equation}
The total charge is defined as:
\begin{equation}\label{1.13}
Q=\int d^2xj^0\ ,
\end{equation}
now taking the form
\begin{equation}\label{1.14}
Q=4\pi\int\rho d\rho\omega{\sigma}^2e^{-\nu/2}e^{\lambda/2}\ .
\end{equation}
This charge represents the total particle number if the charge of
each particle is unity. In the present case the energy of the
free particles with the same charge is $mQ$, where $m$ is the
mass of the free particles. We rescale all the parameters of the
Lagrangian with respect to the mass, and set $m$ eqaul to unity.
So, if the total charge is less than the total mass of the field
configuration, the star is stable with respect to decay into free
particles. We define for convenience:
\begin{equation}\label{1.15}
B=e^{-\nu}\ , \hspace{1em} A=e^{-\lambda}\ ,
\end{equation}
and rescale:
\begin{equation}\label{1.16}
\begin{split}
\tilde{\rho}=\rho m\ , \hspace{1em} \tilde{\omega}=\omega/m\ ,
\hspace{1em} \tilde{\sigma}=\sigma/m^{1/2}\ , \hspace{1em} \\
\widetilde{U}=U/m^3\ , \hspace{1em} \widetilde{W}=W/m^3\ ,
\hspace{1em} \widetilde{V}=V/m^3\ ,
\end{split}
\end{equation}
where $m$ in generally is a mass scale, here the mass of the
scalar field. We also make the rescalings
\begin{equation}\label{1.17}
\widetilde{\Lambda}8\pi Gm^3\equiv\Lambda, \hspace{1em}
\tilde{r}=\sqrt{8\pi Gm}\tilde{\rho}\ .
\end{equation}

Inside the soliton both metric and matter fields vary very slowly
with respect to the radius. This leads to a considerable
simplification to both Einstein and Lagrange equations. We see
that $V\propto{\phi(0)}^2/{P}^2\propto
m^3{\tilde{\sigma}(0)}^2/{\tilde{P}}^2\propto
Gm^4{\tilde{\sigma}}^2/{\tilde{R}}^2$ which is very small compared
to the other energy quantities. $P$ is the (unrescaled) soliton
radius and $\tilde{P}$ and $\tilde{R}$ correspond to the rescaled
radii, according to the rescalings under \ref{1.16} and \ref{1.17}
respectively.

So, ignoring the $O(Gm^4)$ quantities and dropping the tildes the
Einstein equations take the form:
\begin{equation}\label{1.18}
-\frac{dA}{dr}\frac{1}{2r}={\omega}^2{\sigma}^2B+U+\Lambda\ ,
\end{equation}
\begin{equation}\label{1.19}
-\frac{A}{B}\frac{dB}{dr}\frac{1}{2r}={\omega}^2{\sigma}^2B-U-\Lambda\
,
\end{equation}
and the Euler-Lagrange equation is:
\begin{equation}\label{1.20}
{\omega}^2\sigma B-\frac{1}{2}\frac{dU}{d\sigma}=0\ .
\end{equation}
A rescaled potential admitting q-ball type solutions in the
absence of gravity is:
\begin{equation}\label{1.21}
U={|\phi|}^2-{|\phi|}^4+\frac{1}{3}{|\phi|}^6={\sigma}^2
\left(1-{\sigma}^2+\frac{1}{3}{\sigma}^4\right)\ .
\end{equation}
The solution to the eq. \ref{1.20} is:
\begin{equation}\label{1.22}
{\sigma}^2=1+\omega\sqrt{B}
\end{equation}
and $U$ is:
\begin{equation}\label{1.23}
U=\frac{1}{3}(1+{\omega}^3B^{3/2})\ .
\end{equation}

Within the thin surface the metric and matter fields change
rapidly. We can write the Lagrange equation in the following form
(unrescaled quantities):
\begin{equation}\label{1.24}
\left(\frac{A}{\rho}-\frac{1}{2}\frac{A}{B}\frac{dB}{d\rho}
\right)\frac{d\sigma}{d\rho}=\frac{1}{2}
\frac{\delta(U-W-V)}{\delta\sigma}\ .
\end{equation}
There is a certain form for the $g_{11}(r)$ metric component
(unrescaled quantities), \cite{lubo}, \cite{astefanesei}:
\begin{equation}\label{1.25}
-g_{11}^{-1}(\rho)\equiv A(\rho)=1-\frac{2GE_{\rho}}{\rho^{D-3}}-
\frac{2\Lambda\rho^2}{(D-2)(D-1)} ,
\end{equation}
where $E_{\rho}$ is the total energy density contained within a
sphere of radius $\rho$ and $D$ is the spacetime dimensionality,
here equal to $3$. The above relation can be verified numerically
and will be discussed more extensively when dealing with q-stars
with two scalar fields, which offer a richer variety of analytical
solutions. Using the relation holding within the surface
approximately namely the: $A\sim1/B$ and rescaling in the usual
way, we find that:
\begin{equation}\label{1.25a}
\frac{1}{2} \frac{\delta(U-W-V)}{\delta\sigma}\sim \sqrt{8\pi Gm}
\frac{1-\frac{1}{\pi}\int_0^{\tilde{r}}d^2\tilde{r}\tilde{\varepsilon}-
\frac{1}{\pi}\tilde{r}\frac{d}{d\tilde{r}}\int_0^{\tilde{r}}d^2\tilde{r}\tilde{\varepsilon}-
2\widetilde{\Lambda}{\tilde{r}}^2}{\tilde{r}}m^4\ ,
\end{equation}
where we regarded that within the surface the field decreases
from a ${\sigma}_0$ value to zero and that ${\sigma}_0\sim
m^{1/2}$. Also, the surface width is $\sim m^{-1}$. So, dropping
the $O(\sqrt{Gm})$ quantities, we find:
\begin{equation}\label{1.26}
\frac{\delta(W-U-V)}{\delta\sigma}=0\ .
\end{equation}
Because at the outer edge of the surface all energy quantities
are zero, we find a first integral to the above equation:
\begin{equation}\label{1.27}
V+U-W=0\ .
\end{equation}
Eqs. \ref{1.25a}-\ref{1.27} hold true only within the surface. At
the inner edge of the surface $\sigma'$ is zero in order to match
the interior with the surface solution. So, at the inner edge of
the surface the equality $U=W$, together with eqs. \ref{1.22},
\ref{1.23}, gives the eigenvalue equation for $\omega$:
\begin{equation}\label{1.28}
\omega=\frac{B_{\textrm{sur}}^{-1/2}}{2}=\frac{A_{\textrm{sur}}^{1/2}}{2}\
.
\end{equation}
which has the right limiting value when gravity is absent, i.e.:
$A_{sur}\rightarrow 1$. So, ${\omega}$, apart from a certain
parameter referring to the soliton properties, is also a measure
for the gravity strength, equivalent to the metrics. The same
discussion, eqs. \ref{1.24}-\ref{1.28}, holds true in any case of
q-star.

Solving the Einstein equations and in order to find the soliton
parameters, we use the $G^0_{\ 0}$ component of the Einstein
tensor for the calculation of the energy or, equivalently, the
form
$$E=\pi(1-A(r)-\Lambda r^2)\ ,\hspace{1em} r\rightarrow\infty\ ,$$
(rescaled quantities) arising from eq. \ref{1.25}, and eq.
\ref{1.14} for the calculation of the charge. The thin surface
has a negligible energy and charge contribution. The independent
variables are two: The cosmological constant, a global feature of
the spacetime, and the soliton frequency.

Outside the soliton the energy quantities are all zero. Einstein
equations can be solved analytically:
\begin{equation}\label{extA}
A(r)=A_{\textrm{sur}}-\Lambda r^2+\Lambda R^2=
1-\frac{E}{\pi}-\Lambda r^2\ ,
\end{equation}
\begin{equation}\label{extB}
B(r)=\frac{A_{\textrm{sur}}-\Lambda r^2+\Lambda
R^2}{A_{\textrm{sur}}^2}={\left(1-\frac{E}{\pi}-\Lambda
r^2\right)}^{-1}\ ,
\end{equation}
with $R$ the star radius and $E$ the total mass. The same
relations hold true for every case of q-star we discuss.

\begin{figure}
\centering
\includegraphics{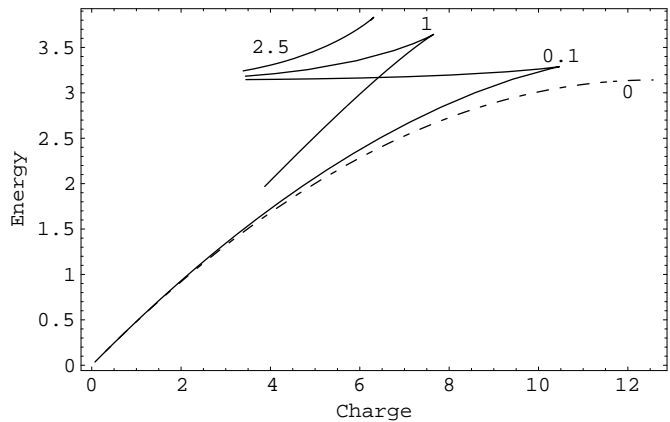}
\caption{Energy as a function of charge for several values of the
cosmological constant for a q-star with one scalar field. The
numbers within the figures \ref{figure1.1}, \ref{figure2.1},
\ref{figure3.1} and \ref{figure4.2}-\ref{figure4.5} denote the
absolute value of the cosmological constant. We start from a large
value of the metric at the surface, $A_{\textrm{sur}}$, near unity,
corresponding to the weak gravity limit . The values of the energy
and charge are small. As we decrease the metric the values of energy
and charge increase. If $\Lambda\neq0$ and when reaching a certain
value of $A_{\textrm{sur}}$, then energy and charge decrease. The
initial increase in the energy and charge values is not visible in
the $\Lambda=-2.5$ case, because it takes place nearly to the
maximum energy and for a short range of the surface gravity
parameter.} \label{figure1.1}
\end{figure}

\begin{figure}
\centering
\includegraphics{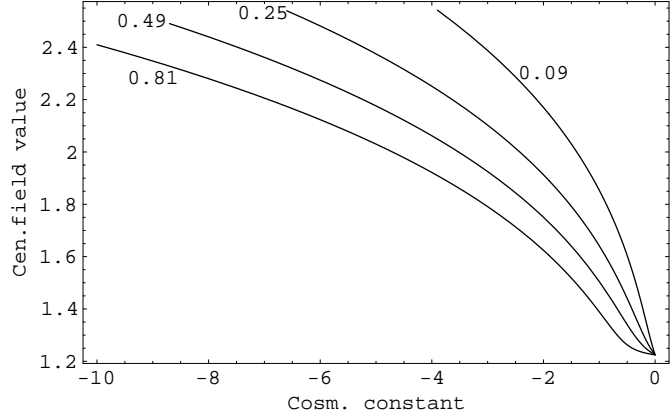}
\caption{The matter field value at the center of the soliton as a
function of the cosmological constant for a q-star with one scalar
field for four different values of $A_{\textrm{sur}}$. The
frequency can be obtained using relation \ref{1.28}. For zero
cosmological constant the field is everywhere constant within the
soliton, equal to
${(1+\omega/A_{\textrm{sur}}^{1/2})}^{1/2}={1.5}^{1/2}$. As the
cosmological constant increases in absolute values the field
takes larger values as a reflection of the increase in the
gravity strength. Also, the field has larger values when $\omega$
decreases, because small $\omega$ indicates stronger gravity.}
\label{figure1.2}
\end{figure}

\begin{figure}
\centering
\includegraphics{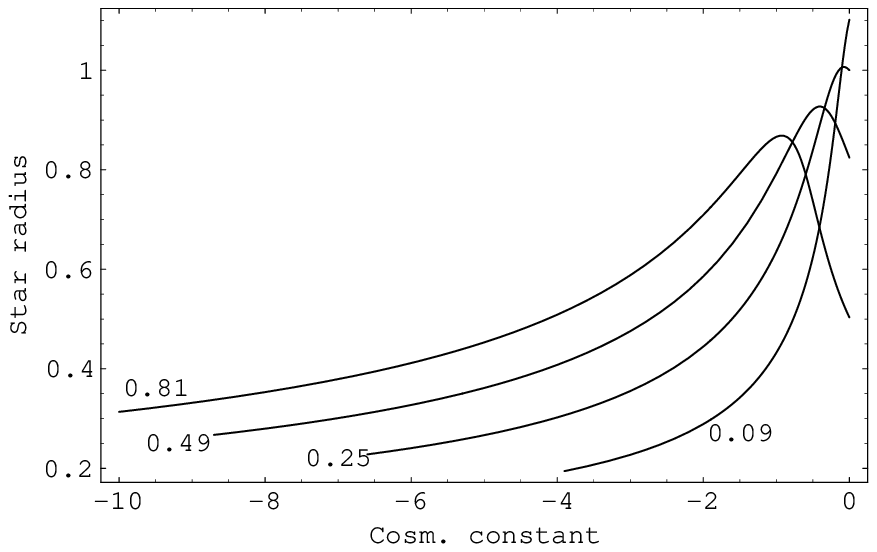}
\caption{The soliton radius as a function of the cosmological
constant for a q-star with one scalar field for four different
values of $A_{\textrm{sur}}$.} \label{figure1.3}
\end{figure}

\begin{figure}
\centering
\includegraphics{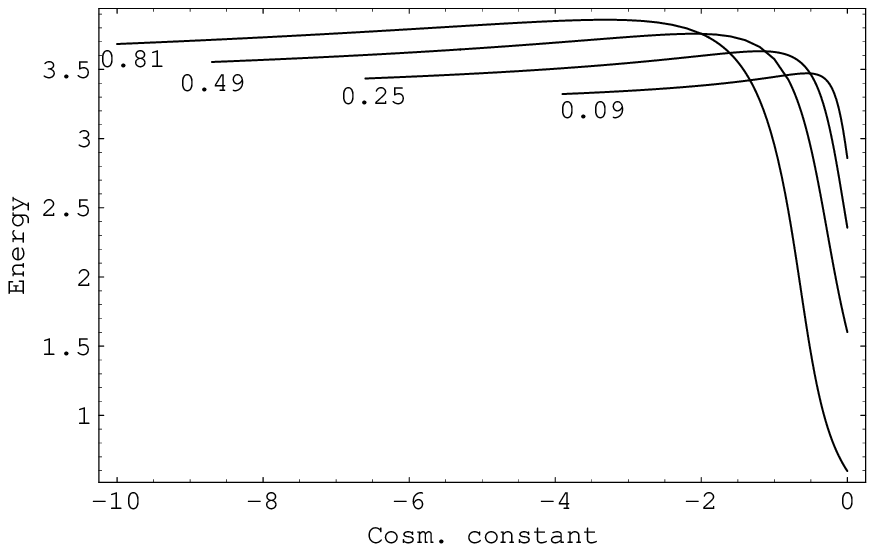}
\caption{The total soliton mass as a function of the cosmological
constant for a q-star with one scalar field for four different
values of $A_{\textrm{sur}}$.} \label{figure1.4}
\end{figure}

\begin{figure}
\centering
\includegraphics{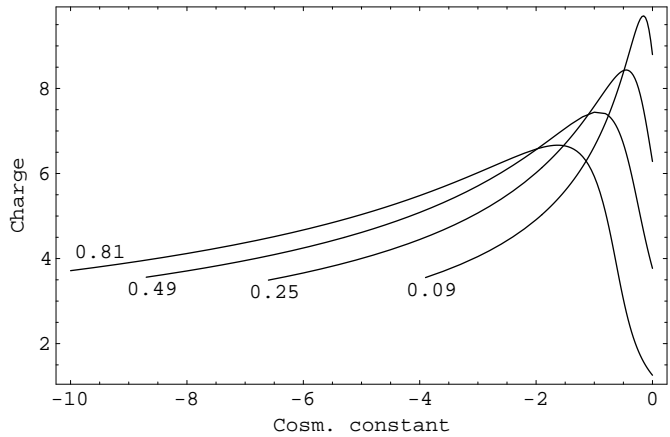}
\caption{The particle number as a function of the cosmological
constant for a q-star with one scalar field for four different
values of $A_{\textrm{sur}}$. We find that radius, energy and
charge increase for small absolute values of the cosmological
constant but for larger values they decrease.} \label{figure1.5}
\end{figure}

We will now give the analytic solution holding true when
$\Lambda\rightarrow 0$. In this case the metric $B$ is constant
everywhere and the other parameters of the soliton are given in
the following equations:
\begin{equation}\label{1.29}
B(r)=1/A_{\textrm{sur}}=\frac{1}{4{\omega}^2}\ ,
\end{equation}
\begin{equation}\label{1.30}
A(r)=1-\frac{3}{4}r^2\ ,
\end{equation}
\begin{equation}\label{1.31}
R=\sqrt{\frac{4}{3}(1-A_{\textrm{sur}})}=\sqrt{\frac{4}{3}(1-4{\omega}^2)}\
,
\end{equation}
\begin{equation}\label{1.32}
E=\pi(1-A_{\textrm{sur}})=\pi(1-4{\omega}^2)\ ,
\end{equation}
\begin{equation}\label{1.33}
Q=4\pi(1-\sqrt{A_{\textrm{sur}}})=4\pi(1-2\omega)\ .
\end{equation}
In the above equations $R$ is the soliton radius determined when
$A(r)=1/B(r)$ and $A_{\textrm{sur}}$ is the value of the $A(r)$ at
the surface of the star. There are some interesting results
obtained when investigating the properties of the analytical
solution. The total soliton energy is:
\begin{equation}\label{1.37}
E=\frac{3\pi}{4}R^2\ .
\end{equation}
This agrees with the intuitive result that the total energy is
analogous to the ``volume" of the field configuration. Also,
$E/Q<1=$ the rescaled mass of the free particles for every
$0<A_{\textrm{sur}}\leq1$ for the case of zero cosmological
constant. This means that there is no possibility for the soliton
fission to free particles, because this procedure is energetically
forbidden. From the $\sigma^2=1+\omega B^{1/2}$ relation and eq.
\ref{1.29} we find that in a flat spacetime, the scalar field has
an everywhere constant absolute value, here equal to $1.5$. So,
when $A_{\textrm{sur}}\rightarrow0$, i.e.: when a horizon is
going to be formed, there is no anomaly at the center of the star.

In our figures, every dashed line corresponds to an analytical
solution. The total energy is in $8\pi G$ units, the particle
number in $8\pi G \tilde{m}$ units, with $\tilde{m}=1$ and the
soliton radius in ${(8\pi G\tilde{m}^3)}^{1/2}$ units. In figures
\ref{figure1.2}-\ref{figure1.5} starting with an initial zero
value of the cosmological constant, we increase its absolute
value and terminate our calculations when the soliton energy
equates to the energy of free particles with the same charge. In
figure \ref{figure1.1} we start from a large value of the metric
field $A_{\textrm{sur}}$, near unity, or, equivalently, a large
value of the characteristic soliton frequency, near $1/2$ (which
according to eq. \ref{1.28} is the maximum limiting value of the
frequency when gravity is absent, i.e., when
$A_{\textrm{sur}}\rightarrow 1$) and gradually decrease the
frequency.

We can observe, as a general comment of our numerical results
depicted in figures \ref{figure1.1}-\ref{figure1.5}, that the
effect of the increase, in absolute values, of the cosmological
constant causes a consequent increase in the energy, charge and
radius of the q-star. This is a counter-balancing effect: Negative
cosmological constant means that different particles within the
soliton, corresponding to the $\phi$ field, tend to move along
deviating geodesics. So, the increase in mass and the other
soliton parameters is necessary in order the object to avoid this
kind of ``deviation", or, in other words, to generate ``positive"
gravity as opposed to the ``negative" gravity, generated by the
negative cosmological constant. But when cosmological constant
becomes considerably large (i.e. when $\Lambda$ is larger than
$8\pi GU$ or $8\pi GW$ or, equivalently, when
$\widetilde{\Lambda}$ is larger than $\widetilde{U}$ or
$\widetilde{W}$) then no soliton energy contribution can prevent
this ``deviation", apart from a shrinking in the soliton magnitude
and a consequent decrease in the soliton mass and particle number.

Figure \ref{figure1.1} gives more interesting results. We start
from an $\omega$ near the critical value $1/2$ holding true in
the absence of gravity. The field configuration is a small one,
with small energy and charge. When decreasing $\omega$ (or
$A_{\textrm{sur}}$) gravity becomes more important and the
soliton becomes larger and more massive. There is an interesting
point, common in similar diagrams of other kinds of q-stars.
Decreasing frequency both energy and charge increase, but only
until the frequency reaches a certain point, below which the two
quantities decrease. So for a certain charge there are usually
two values of the energy, the lower one corresponding to high
frequency (lower branch of the $E=E(Q)$ diagram) and the higher
one corresponding to low frequency (upper branch of the $E=E(Q)$
diagram). So, a soliton star with a certain charge and a high
energy amount can emit the energy excess, falling down to the
lower branch, changing of course its frequency. We will study
this case in more details when treating fermion-scalar q-stars,
for which there is an analytical solution for a special
$\Lambda\neq0$ case and the obtained relations are simpler than
in the case of q-stars with two scalar fields. This change in the
energy and charge variation with respect to the frequency happens
only when the cosmological constant differs from zero. The above
analytical solution given for the zero cosmological constant case
predicts that both energy and charge are monotone decreasing
functions of the frequency. When $A_{\textrm{sur}}\rightarrow0$
one expects that a black hole is formed. But the calculations
depicted in figure \ref{figure1.1} interrupted for an
$A_{\textrm{sur}}>0$ value, because below that certain value the
star decays into free particles as the energetically favorable
choice. This means that before the formation of an horizon, the
q-star decays. So, when $\Lambda\neq0$ gravitational collapse is
impossible, at least in the usual way of decreasing frequency
holding true for q-stars in $3+1$ dimensions,
\cite{lyn1}-\cite{lyn5}. There may be a possibility of forming
q-type black holes, if one takes into account the spatial
variation of the scalar field, but this demands a considerably
different way of solving the equations of motion to the one known
so far. The same discussion holds true for any sort of q-star.

\initiate
\section{A q-star with two scalar fields}

A first simple generalization to the above described soliton is a
field configuration with two scalar fields, one N-carrying $\phi$,
and, consequently, complex, and the other, $\sigma$, used to
constrain the N-carrying field within a certain region,
generating an appropriate potential, taken to be real for
simplicity. The Lagrangian in the case under discussion is:
\begin{equation}\label{2.1}
L=g^{\mu\nu}{({\partial}_{\mu}\phi)}^{\ast}({\partial}_{\nu}\phi)
+\frac{1}{2}g^{\mu\nu}({\partial}_{\mu}\sigma)({\partial}_{\nu}\sigma)-U\
,
\end{equation}
and the metric is supposed to have the same form as in the q-star
with one scalar field, i.e. to be static and spherically
symmetric. The potential has the general form:
\begin{equation}\label{2.2}
U=a{\phi}^2{\sigma}^2+b{\phi}^4+c{({\sigma}^2-d)}^2\ .
\end{equation}
We paramertize $d$ as:
\begin{equation}\label{2.3}
d={\mu}^{-3/2}a^{-1/2}b^{1/2}\ ,
\end{equation}
where the parameter $\mu$ has mass dimensions, is nothing more
than a way of re-expressing $d$, is of the same order of
magnitude as the potential parameters and will simplify the
rescalings of the energy quantities. We now rescale the matter
fields
\begin{equation}\label{2.4}
\tilde{\phi}={(2b)}^{1/4}{\mu}^{-4/3}\phi, \hspace{1em}
\tilde{\sigma}=a^{1/2}{(2b)}^{-1/4}{\mu}^{3/4}\sigma\ .
\end{equation}
We also define:
\begin{equation}\label{2.5}
\lambda=a^{-1}bc\ ,
\end{equation}
and the potential takes the simple form:
\begin{equation}\label{2.6}
U=\widetilde{U}/{\mu}^3\ , \hspace{1em}
\widetilde{U}={\tilde{\phi}}^2{\tilde{\sigma}}^2+\frac{1}{2}{\tilde{\phi}}^4+
\frac{\lambda}{2}{({\tilde{\sigma}}^2-1)}^2\ .
\end{equation}
The rescalings in the other quantities (spacetime and frequency)
are the same as in eq. \ref{1.16}.

We use a simple ansatz for the matter fields, namely we regard
that within the soliton holds:
\begin{equation}\label{2.7}
\tilde{\phi}(\tilde{\vec{\rho}},t)=\tilde{\varphi}(\tilde{\rho})e^{-\imath\tilde{\omega}t}\
, \hspace{1em} \tilde{\sigma}\cong0\ .
\end{equation}
Redefining: $\tilde{r}=\tilde{\rho}8\pi G\mu$, we have:
\begin{equation}\label{2.8}
\widetilde{V}=
A\left[{\left(\frac{d\tilde{\varphi}}{d\tilde{\rho}}\right)}^2
+{\left(\frac{d\tilde{\sigma}}{d\tilde{\rho}}\right)}^2\right]
\cong {\left(\frac{d\tilde{\varphi}}{d\tilde{\rho}}\right)}^2
\propto {\left(\frac{d\tilde{\varphi}}{d\tilde{r}}\right)}^2 8\pi
G\mu \ll \widetilde{W}\ , \widetilde{U}\ ,
\end{equation}
with
\begin{equation}\label{2.9}
\widetilde{W}=B{\tilde{\omega}}^2{\tilde{\varphi}}^2\ .
\end{equation}

The Euler-Lagrange equation for the complex field is:
\begin{equation}\label{2.10}
A\left[\frac{d^2\tilde{\varphi}}{d\tilde{r}^2}+\left(\frac{1}{\tilde{r}}+\frac{1}{2}
\frac{d(A-B)}{d\tilde{r}}\right)\frac{d\tilde{\varphi}}{d\tilde{r}}\right]8\pi
G\mu =\frac{1}{2}
\frac{\delta(\widetilde{U}-\widetilde{W})}{\delta\tilde{\varphi}}\
.
\end{equation}
Regarding the metric and matter fields as smoothly varying within
the star, we find from the Lagrange equation for the $\varphi$
field:
\begin{equation}\label{2.11}
{\tilde{\varphi}}^2={\tilde{\omega}}^2B\ ,
\end{equation}
\begin{equation}\label{2.12}
\begin{split}
\widetilde{U}=\frac{1}{2}({\tilde{\omega}}^4B^2+\lambda) \\
\widetilde{W}={\tilde{\omega}}^4B^2\ .
\end{split}
\end{equation}
Within the surface the metric and matter fields vary rapidly.
Repeating the discussion (eqs \ref{1.24}-\ref{1.27}) of the
previous section we find the eigenvalue equation for the
frequency:
\begin{equation}\label{2.13}
\tilde{\omega}={\left(\frac{\lambda}{B_{\textrm{sur}}^2}\right)}^{1/4}\
.
\end{equation}
We will choose $\lambda=1/9$ for simplicity. Dropping from now on
the tildes, the Einstein equations take the form:
\begin{equation}\label{2.14}
-\frac{1}{2r}\frac{dA}{dr}=\frac{3}{2}{\omega}^4B^2+\frac{1}{2}\lambda+
\Lambda\ ,
\end{equation}
\begin{equation}\label{2.15}
-\frac{1}{2r}\frac{A}{B}\frac{dB}{dr}=\frac{1}{2}{\omega}^4B^2-\frac{1}{2}\lambda-\Lambda\
.
\end{equation}
The soliton mass is given by the $G^0_{\ 0}$ component of the
Einstein tensor and the particle number is:
\begin{equation}\label{2.16}
Q=4\pi\int drr\omega{\varphi}^2\sqrt{\frac{B}{A}}= 4\pi\int
drr{\omega}^3B^{3/2}A^{-1/2}\ .
\end{equation}

Outside the soliton we take $\varphi$ to be everywhere zero and
$\sigma$ equal to unity, so that both the energy density and
Noether current are zero. These values are also solutions to the
Lagrange equations of the matter fields.

\begin{figure}
\centering
\includegraphics{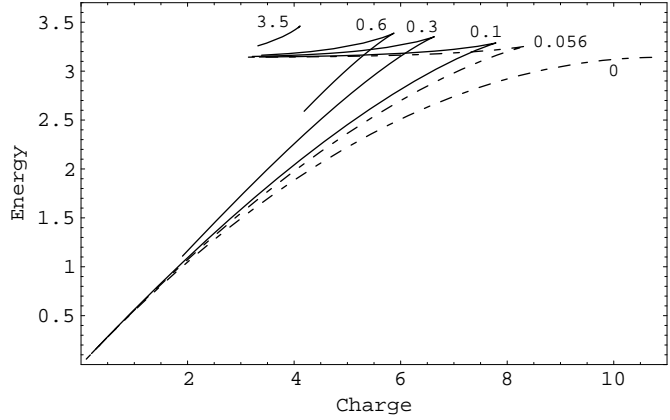}
\caption{The total mass of the field configuration as a function
of the particle number for several values of the cosmological
constant for a q-star with two scalar fields. Dashed lines depict
analytical solutions.} \label{figure2.1}
\end{figure}

\begin{figure}
\centering
\includegraphics{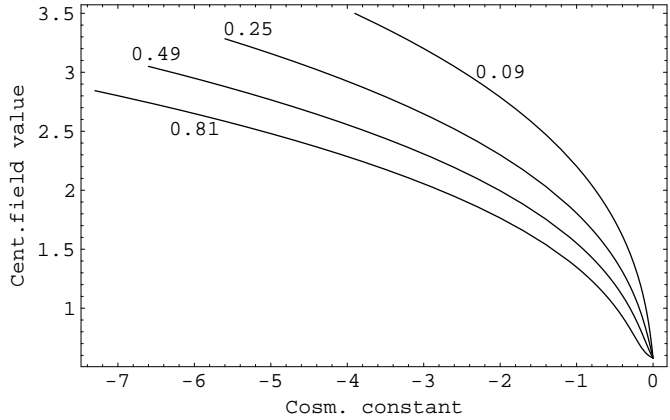}
\caption{The absolute value of the $N-$carrying field at the
center of the field configuration as a function of the
cosmological constant for a q-star with two scalar fields for
four values of $A_{\textrm{sur}}$, or equivalently, the
frequency. For flat spacetime the field $\phi$ has a constant
value equal to ${\lambda}^{1/4}$ (here ${1/9}^{1/4}$) in the
soliton interior.} \label{figure2.2}
\end{figure}

\begin{figure}
\centering
\includegraphics{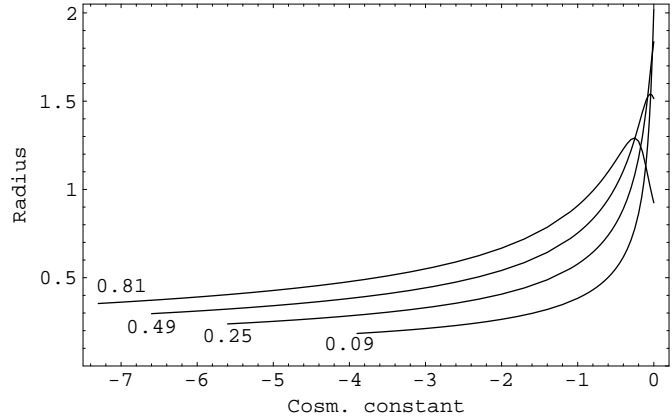}
\caption{The radius of a q-star with two scalar fields as a
function of the cosmological constant for four values of
$A_{\textrm{sur}}$.} \label{figure2.3}
\end{figure}

\begin{figure}
\centering
\includegraphics{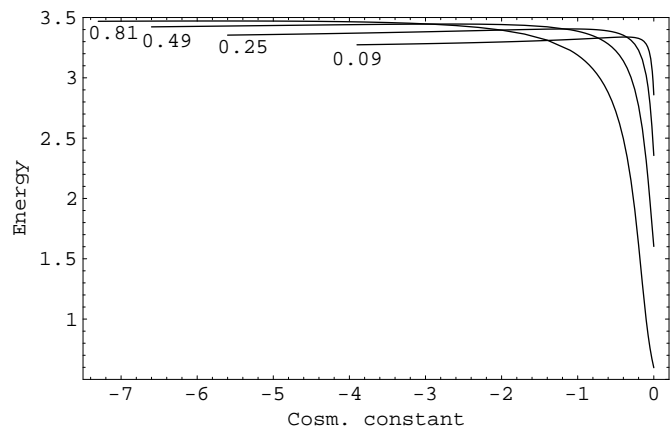}
\caption{The total mass as a function of the cosmological
constant for a q-star with two scalar fields for four values of
$A_{\textrm{sur}}$.} \label{figure2.4}
\end{figure}

\begin{figure}
\centering
\includegraphics{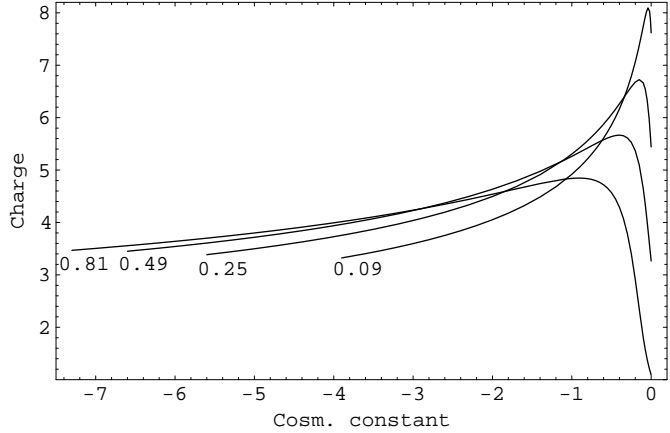}
\caption{The particle number as a function of the cosmological
constant for a q-star with two scalar fields for four values of
the $A_{\textrm{sur}}$.} \label{figure2.5}
\end{figure}

We will now find some analytical solutions. The case of zero
cosmological constant is one of them. In this case holds:
\begin{equation}\label{2.17}
B(r)=1/A_{\textrm{sur}}\ , \hspace{1em} A(r)=1-2\lambda r^2\ .
\end{equation}
The soliton radius $R$ can be found equal to:
\begin{equation}\label{2.18}
R={\left(\frac{1-A_{\textrm{sur}}}{2\lambda}\right)}^{1/2}=
{\left(\frac{1-{\omega}^2/{\lambda}^{1/2}}{2\lambda}\right)}^{1/2}\
,
\end{equation}
and the energy and charge
\begin{equation}\label{2.19}
E=(1-A_{\textrm{sur}})\pi=(1-{\omega}^2/{\lambda}^{1/2})\pi\ ,
\end{equation}
\begin{equation}\label{2.20}
Q=2\sqrt{3}\pi(1-A_{\textrm{sur}}^{1/2})=2\sqrt{3}\pi(1-\omega/{\lambda}^{1/4})\
.
\end{equation}
We can see that the equation $E=Q$ has no solution, so, the
soliton energy can not be equal to the energy of free particles
and consequently  the star can not decay into free particles. We
can also find the relation for the energy $E_r$ stored within a
surface of radius $r$:
\begin{equation}\label{2.20a}
E_r=2\lambda\pi r^2\ ,
\end{equation}
which verifies relation \ref{1.25}, when combined with eq.
\ref{2.17}.

We will now examine a different case admitting analytical
solutions. The case:
\begin{equation}\label{2.21}
\frac{1}{2}\lambda+\Lambda=0\ .
\end{equation}
If $R$ is the soliton radius (i.e. the solution to the equation:
$A(r)=1/B(r)$) the soliton parameters take the form:
\begin{equation}\label{2.22}
R=3\sqrt{2}{(A_{\textrm{sur}}^{2/3}-A_{\textrm{sur}})}^{1/2}=
3\sqrt{2}{(-3{\omega}^2+3^{2/3}{\omega}^{4/3})}^{1/2}\ ,
\end{equation}
\begin{equation}\label{2.23}
A(r)=\frac{{(18A_{\textrm{sur}}-r^2+R^2)}^3}{5832A_{\textrm{sur}}^2}=
\frac{{(18A_{\textrm{sur}}^{2/3}-r^2)}^3}{5832A_{\textrm{sur}}^2}=
\frac{{(-r^2+18\cdot3^{2/3}{\omega}^{4/3})}^3}{52488{\omega}^4}\ ,
\end{equation}
\begin{equation}\label{2.24}
B(r)=\frac{18A_{\textrm{sur}}-r^2+R^2}{18A_{\textrm{sur}}^2}=
\frac{18A_{\textrm{sur}}^{2/3}-r^2}{18A_{\textrm{sur}}^2}=
\frac{-r^2+18\cdot3^{2/3}{\omega}^{4/3}}{162{\omega}^4}\ ,
\end{equation}
\begin{equation}\label{2.25}
E=(1+A_{\textrm{sur}}^{2/3}-2A_{\textrm{sur}})\pi=
(1-6{\omega}^2+3^{2/3}{\omega}^{4/3})\pi\ ,
\end{equation}
\begin{equation}\label{2.26}
Q=4\sqrt{3}\pi(A_{\textrm{sur}}^{1/6}-A_{\textrm{sur}}^{1/2})=
4\sqrt{3}\pi(3^{1/6}{\omega}^{1/3}-3^{1/2}\omega)\ .
\end{equation}
The equation $E=Q$ admits analytical solution numerically equal
to $A_{\textrm{sur}}=1.03484\cdot 10^{-5}$. Then, $E\cong
Q\cong\pi$ and below this value of the surface metric the energy
of the free particles with the same charge is less than the
soliton energy, making in this way stars with $A_{\textrm{sur}}$
below that critical value unstable. We can also find $E_r$:
\begin{equation}\label{2.26a}
E_r=\frac{\pi
r^2(792A_{\textrm{sur}}^{4/3}+324A_{\textrm{sur}}^2-54A_{\textrm{sur}}^{2/3}r^2+
r^4)}{5832A_{\textrm{sur}}^2}\ .
\end{equation}
It is a matter of simple algebra to verify equation \ref{1.25}.

As one can see from figure \ref{figure2.1}, the analytical
solutions behave in the same way as the solutions obtained in
numerical methods. Summarizing our results, we see that decay
into free particles is forbidden for the flat spacetime case,
energetically favorable for an anti de Sitter spacetime, but only
below a certain value of the frequency. Solitons can not be
extremely large. For any value of the independent parameters
($\omega$ and $\Lambda$) there is a certain region in the $E$-$Q$
phase space, fully depicted in our figures,
\ref{figure2.1}-\ref{figure2.5}.

\initiate
\section{Non-abelian q-stars}

The Lagrangian in this case is:
\begin{equation}\label{3.1}
\mathcal{L}=\frac{1}{2}\textrm{Tr}({\partial}_{\mu}\phi)({\partial}_{\nu}\phi)-
\textrm{Tr}U(\phi)\ .
\end{equation}
The Enstein equations are:
\begin{equation}\label{3.2}
-\frac{1}{2\rho}\frac{dA}{d\rho}=8\pi
G\left[U+\textrm{Tr}\frac{1}{2}B{\left(\frac{\partial\phi}{\partial
t}\right)}^2+\textrm{Tr}\frac{1}{2}A{\left(\frac{\partial\phi}{\partial\rho}\right)}^2\right]
+\Lambda\ ,
\end{equation}
\begin{equation}\label{3.3}
-\frac{1}{2\rho}\frac{A}{B}\frac{dB}{d\rho}=8\pi
G\left[\textrm{Tr}\frac{1}{2}B{\left(\frac{\partial\phi}{\partial
t}\right)}^2+
\textrm{Tr}\frac{1}{2}A{\left(\frac{\partial\phi}{\partial\rho}\right)}^2-U\right]-\Lambda\
,
\end{equation}
with a general renormalizable potential
\begin{equation}\label{3.4}
U=\frac{{\mu}^2}{2}{\phi}^2+\frac{g}{3!}{\phi}^3+\frac{\lambda}{4!}{\phi}^4\
,
\end{equation}
with $\phi$ in the $SO(3)$ \textbf{5} representation. We make the
following rescalings:
\begin{equation}\label{3.5}
g=\mu\tilde{g}\ , \hspace{0.2cm}
\phi=(\mu/\tilde{g})\tilde{\phi}\ , \hspace{0.2cm}
\lambda={\tilde{g}}^2\tilde{\lambda}\ , \hspace{0.2cm}
\rho=\tilde{\rho}{\mu}^{-1}\ , \hspace{0.2cm}
\omega=\mu\tilde{\omega}\ .
\end{equation}
All energy quantities are rescaled by ${\mu}^4/{\tilde{g}}^2$.
The potential takes the simple form:
\begin{equation}\label{3.6}
U=\frac{{\mu}^4}{{\tilde{g}}^2}\widetilde{U}\ , \hspace{1em}
\widetilde{U}=\frac{{\tilde{\phi}}^2}{2}+\frac{{\tilde{\phi}}^3}{3!}+
\frac{\tilde{\lambda}}{4!} {\tilde{\phi}}^4\ ,
\end{equation}
where $\tilde{\lambda}=\lambda/{\tilde{g}}^2$ Redefining
$$\tilde{r}=\tilde{\rho}\sqrt{8\pi
G\frac{{\mu}^2}{{\tilde{g}}^2}}\ ,$$ $$8\pi
G\frac{{\mu}^4}{{\tilde{g}}^2}\widetilde{\Lambda}=\Lambda\ ,$$ the
independent Einstein equations take the simple form:
\begin{equation}\label{3.7}
-\frac{1}{2\tilde{r}}\frac{dA}{d\tilde{r}}=\mathcal{U}+\mathcal{W}
+\mathcal{V}+\widetilde{\Lambda}\ ,
\end{equation}
\begin{equation}\label{3.8}
-\frac{1}{2\tilde{r}}\frac{A}{B}\frac{dB}{d\tilde{r}}=\mathcal{W}+\mathcal{V}
-\mathcal{U}-\widetilde{\Lambda}\ ,
\end{equation}
where:
\begin{equation}\label{3.9}
\centering
\begin{split}
\mathcal{U}&=\textrm{Tr}\widetilde{U}\ , \\
\mathcal{W}&=\textrm{Tr}\frac{1}{2}B{\left(\frac{\partial\tilde{\phi}}
{\partial\tilde{t}}\right)}^2
=\textrm{Tr}B{[\widetilde{\Omega},\tilde{\phi}]}^2\ , \\
\mathcal{V}&=\textrm{Tr}\frac{1}{2}A{\left(\frac{\partial\tilde{\phi}}
{\partial\tilde{\rho}}\right)}^2=\textrm{Tr}\frac{1}{2}A{\left(\frac{\partial\tilde{\phi}}
{\partial\tilde{r}}\right)}^2 8\pi G
\frac{{\mu}^2}{{\tilde{g}}^2}\ .
\end{split}
\end{equation}
It is obvious that for a large soliton with smoothly varying
fields within the interior the $V$ terms are of $O(G)$ order and,
thus, can be neglected. We define $\Omega$ through:
\begin{equation}\label{3.10}
\frac{\partial\phi}{\partial t}=\imath[\Omega,\phi]\ .
\end{equation}
We also define:
\begin{equation}\label{3.11}
\Omega\equiv\mu\ \widetilde{\Omega}\equiv-\imath\tilde{\omega} \mu
\left(
\begin{array}{ccc}
  0 & 0 & 1 \\
  0 & 0 & 0 \\
  -1 & 0 & 0
\end{array} \right)\ .
\end{equation}
With the above definitions, the Euler-Lagrange equation takes the
form:
\begin{eqnarray}\label{3.12}
-B[\widetilde{\Omega},[\widetilde{\Omega},\tilde{\phi}]]+
\frac{\partial\mathcal{U}}{\partial\tilde{\phi}}-
\frac{1}{3}\mathbf{1}\textrm{Tr}\left(\frac{\partial\mathcal{U}}{\partial\tilde{\phi}}\right)=
\nonumber\\ 8\pi
G\frac{{\mu}^2}{{\tilde{g}}^2}\left[\frac{{\partial}^2\tilde{\phi}}{\partial{\tilde{r}}^2}
+\frac{\partial\tilde{\phi}}{\partial{\tilde{r}}}\left(\frac{1}{\tilde{r}}+
\frac{1}{2A}\frac{dA}{d\tilde{r}}-\frac{1}{2A}\frac{dA}{d\tilde{r}}\right)\right]\
,
\end{eqnarray}
where $\mathbf{1}$ is the unity matrix. We can always diagonalize
$\phi=e^{\imath R}{\phi}_{\textrm{diag}}e^{-\imath R}$. The rigid
rotation condition means that $R(\rho,t)=\Omega t+C$ where we can
eliminate the constant $C$ through a global $SO(3)$ rotation and
write:
\begin{equation}\label{3.13}
\phi=-\frac{1}{2}{\phi}_2\cdot\textrm{diag}(1+y,-2,1-y)\Leftrightarrow
\tilde{\phi}=-\frac{1}{2}{\tilde{\phi}}_2\cdot\textrm{diag}(1+y,-2,1-y)\
.
\end{equation}
The charge is defined by the equation:
\begin{equation}\label{3.14}
Q=\int d^2\rho\sqrt{-g}j^0=-\imath\int
d^2\rho\sqrt{-g}B[\phi,\dot{\phi}]\ .
\end{equation}

The equation of motion for the matter field within the soliton
dropping the $O(8\pi G{\mu}^2/{\tilde{g}}^2)$ takes the form:
\begin{equation}\label{3.15}
\tilde{\phi}+\frac{1}{2}{\tilde{\phi}}^2+
\frac{1}{6}\tilde{\lambda}{\tilde{\phi}}^3-\frac{1}{3}\mathbf{1}
\textrm{Tr}\left(\frac{1}{2}{\tilde{\phi}}^2+\frac{1}{6}{\tilde{\phi}}^3\right)=
2{\tilde{\omega}}^2B(\tilde{{\phi}_1}-{\tilde{\phi}}_3)\textrm{diag}(1,0,-1)\
.
\end{equation}
At the inner edge of the surface we obtain the usual relation
(repeating eqs \ref{1.24}-\ref{1.27}):
\begin{equation}\label{3.16}
{(\mathcal{U}-\mathcal{W})}_{\textrm{sur}}=0\ .
\end{equation}
The usual relation $\delta(W=U)/\delta\phi=0$ is no longer valid.
We can see that that neglecting the $O(8\pi
G{\mu}^2/{\tilde{g}}^2)$ in the Euler-Lagrange equation,
multiplying by $\tilde{\phi}$ and tracing yields
$2\mathcal{W}=\textrm{Tr}(\tilde{\phi}\cdot\partial\mathcal{U}/\partial\tilde{\phi})$.
Applying equation \ref{3.16} we obtain for the fields at the inner
edge of the surface, with $R$ the soliton radius:
\begin{equation}\label{3.17}
{\tilde{\phi}}_{2}(R)=\frac{12}{\tilde{\lambda}}
{\left(\frac{y^2-1}{{(y^2+3)}^2}\right)}_{\textrm{sur}}\ ,
\end{equation}
\begin{equation}\label{3.18}
{\tilde{\omega}}^2B(R)={\left[\frac{1}{4}\left(1+\frac{3}{y^2}\right)-
\frac{3}{4\tilde{\lambda}}\frac{1}{y^2}
\frac{{(y^2-1)}^2}{{(y^2+3)}^2}\right]}_{\textrm{sur}}\ .
\end{equation}
Taking the $(2,2)$ component of the Euler-Lagrange, eq.
\ref{3.15} we take:
\begin{equation}\label{3.19}
y^2(\tilde{r})=\frac{3}{{\tilde{\phi}}_2}
\frac{\tilde{\lambda}{\tilde{\phi}}^2_2+2{\tilde{\phi}}_2+8}
{2-\tilde{\lambda}{\tilde{\phi}}_2}\ .
\end{equation}
Using the above equation combined with the
$2\mathcal{W}=\textrm{Tr}(\tilde{\phi}\cdot\partial\mathcal{U}/\partial\tilde{\phi})$
relation, we find for the ${\tilde{\phi}}_2$ field:
\begin{equation}\label{3.20}
{\tilde{\phi}}_2(\tilde{r})=
\frac{1-4\tilde{\lambda}{\tilde{\omega}}^2B(\tilde{r})}{2\tilde{\lambda}}+
\frac{1}{2\tilde{\lambda}}
{\left({[1-4\tilde{\lambda}{\tilde{\omega}}^2B(\tilde{r})]}^2+
8\tilde{\lambda}[4{\tilde{\omega}}^2B(\tilde{r})-1]\right)}^{1/2}\
.
\end{equation}
Using now eqs \ref{3.17}, \ref{3.19} we find:
\begin{equation}\label{3.21}
{[(\tilde{\lambda}-1){(y^2+3)}^3+16{(y^2+3)}^2-72(y^2+3)+96]}_{\textrm{sur}}=0\
.
\end{equation}
$\tilde{\lambda}<1$ holds in order \ref{3.19} to have real
solutions. The solution to the above equation is:
\begin{eqnarray}\label{3.22}
y^2(R)=-3+\frac{16}{3(1-\tilde{\lambda})}+ \nonumber\\
\frac{2}{3}\frac{2^{1/3}}{-1+\tilde{\lambda}}
{\left[-13+162\tilde{\lambda}-81{\tilde{\lambda}}^2+9{(-1+\tilde{\lambda})}^2
\sqrt{\frac{-1+81{\tilde{\lambda}}^2}{{(-1+\tilde{\lambda})}^2}}\right]}^{1/3}+
\nonumber\\ +\frac{2}{3}\frac{2^{1/3}}{1-\tilde{\lambda}}
{\left[13+162\tilde{\lambda}+81{\tilde{\lambda}}^2+9{(-1+\tilde{\lambda})}^2
\sqrt{\frac{-1+81{\tilde{\lambda}}^2}{{(-1+\tilde{\lambda})}^2}}\right]}^{1/3}\
.
\end{eqnarray}
The idea for the non-abelian solitons is the following: We give a
certain value to the free parameter $\tilde{\lambda}$, in our
calculations $\tilde{\lambda}=2/3$, and using eq. \ref{3.22} we
find $y^2(R)$. Substituting in eq. \ref{3.18} we find the
eigenvalue for the frequency, in our case equal to
$0.4956A_{\textrm{sur}}^{1/2}$. We then substitute the values of
the ${\tilde{\phi}}_2$ and $y_2$ matter fields (eqs
\ref{3.19}-\ref{3.20}) in Einstein equations and solve them
numerically.

\begin{figure}
\centering
\includegraphics{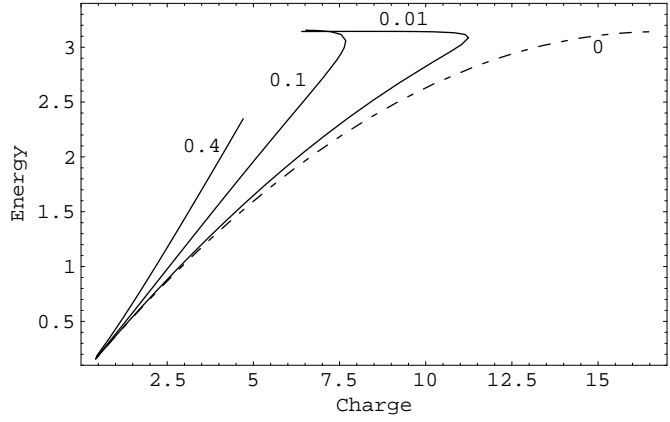}
\caption{Energy as a function of the particle number for four
values of the cosmological constant, for a non-abelian q-star.}
\label{figure3.1}
\end{figure}

\begin{figure}
\centering
\includegraphics{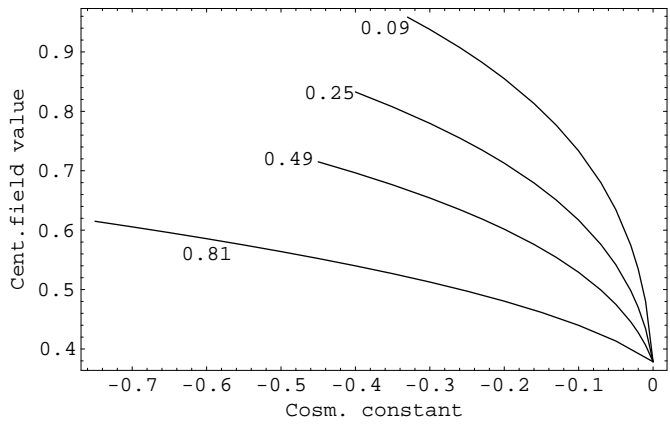}
\caption{The matter field value ($\tilde{\phi}_2(0)$) at the
center of the field configuration as a function of the
cosmological constant for a non-abelian q-star for four values of
the $A_{\textrm{sur}}$.} \label{figure3.2}
\end{figure}

\begin{figure}
\centering
\includegraphics{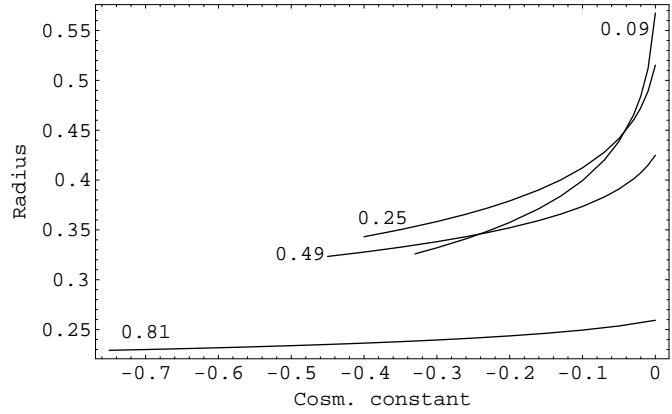}
\caption{The soliton radius as a function of the cosmological
constant for a non-abelian q-star for four values of the
$A_{\textrm{sur}}$.} \label{figure3.3}
\end{figure}

\begin{figure}
\centering
\includegraphics{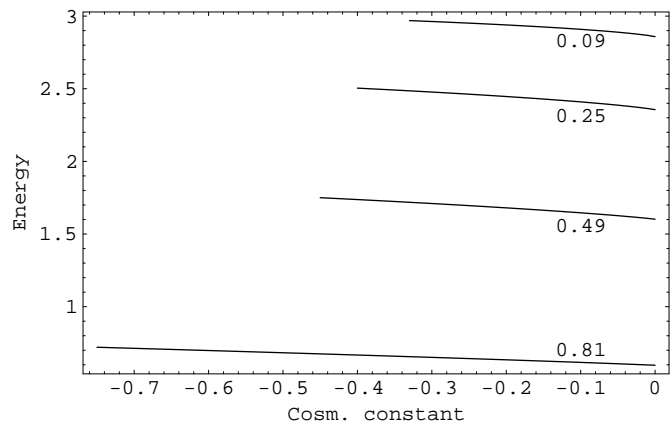}
\caption{The total mass as a function of the cosmological
constant for a non-abelian q-star for four values of the
$A_{\textrm{sur}}$.} \label{figure3.4}
\end{figure}

\begin{figure}
\centering
\includegraphics{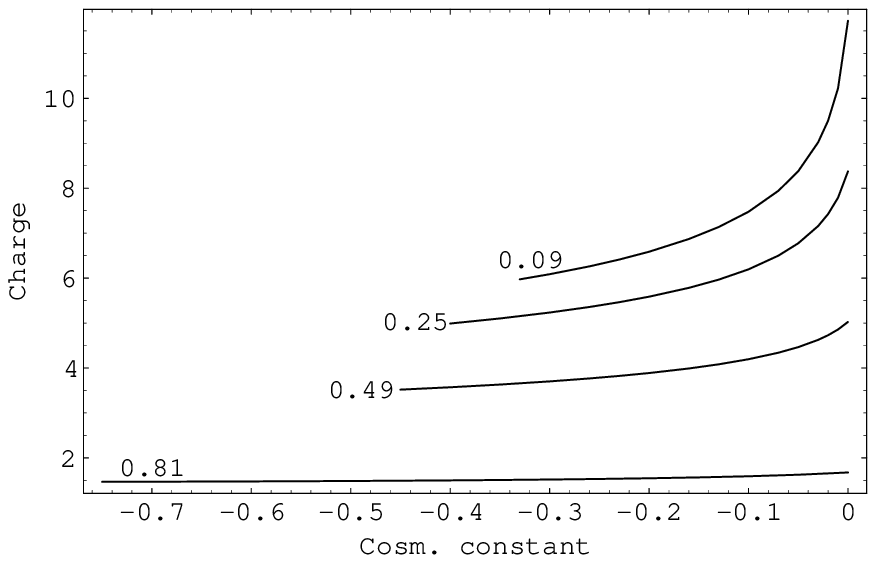}
\caption{The particle number as a function of the cosmological
constant for a non-abelian q-star for four values of the
$A_{\textrm{sur}}$.} \label{figure3.5}
\end{figure}

From the figures \ref{figure3.1}-\ref{figure3.5} we see that for
larger values of the cosmological constant the matter field
($\tilde{\phi}_2$) is larger, but the soliton radius, charge and
energy decrease with the increase of the cosmological constant in
absolute values as in other kinds of q-stars.

We now investigate the analytical solution to the Einstein
equations in the case of zero cosmological constant. We find:
\begin{equation}\label{3.23}
A=1-2.82529{\tilde{r}}^2\ ,
\end{equation}
\begin{equation}\label{3.24}
R={[(1-A_{\textrm{sur}})/2.8259]}^{1/2}\ ,
\end{equation}
\begin{equation}\label{3.25}
E=\pi(1-A_{\textrm{sur}})\ ,
\end{equation}
\begin{equation}\label{3.26}
Q=16.7482(1-A_{\textrm{sur}}^{1/2})\ .
\end{equation}

\initiate
\section{Fermion-Scalar q-stars}

Fermion-scalar q-stars are realistic field configurations of a
scalar and a fermionic field. These field configurations are
supposed, \cite{lyn3}, to describe certain stellar objects known
as neutron stars. The fermionic field carries the charge needed to
stabilize the soliton and the real scalar field generates an
everywhere positive potential. In the star interior the scalar
field is approximately zero but outside the soliton takes a
certain value ${\sigma}_0$, vanishing the potential. The
potential energy density outside the soliton vanishes in this way
(eq. \ref{4.9}) and the fermion mass acquires its vacuum value
$m$.

We regard the fermion-scalar q-star as a zero temperature
fermionic sea with local Fermi energy and momentum
${\varepsilon}_F^2+k_F^2+m^2(\sigma)$. With spherical symmetry,
the local fermion density is:
\begin{equation}\label{4.1}
\langle{\psi}^{\dagger}\psi\rangle=\frac{1}{4{\pi}^2}\int n_kd^2k
=\frac{k_F^2}{4\pi}\ ,
\end{equation}
and the scalar density:
\begin{equation}\label{4.2}
\langle\bar{\psi}\psi\rangle=\frac{1}{4{\pi}^2}\int
n_kd^2k\frac{m}{2{(k^2+m^2)}^{1/2}}=\frac{m(-m+{\varepsilon}_F)}{2\pi}\
.
\end{equation}
We also find using the definitions for the energy and pressure
density:
\begin{equation}\label{4.3}
P_{\psi}\equiv\frac{1}{2}\int
n_kd^2k\frac{k^2}{2{(k^2+m^2)}^{1/2}}=
\frac{1}{3}({\varepsilon}_F\langle{\psi}^{\dagger}\psi\rangle
-m\langle\bar{\psi}\psi\rangle)\ ,
\end{equation}
\begin{equation}\label{4.4}
{\mathcal{E}}_{\psi}\equiv\frac{1}{4{\pi}^2}\int
n_kd^2k{(k^2+m^2)}^{1/2}=2P_{\psi}+m\langle\bar{\psi}\psi\rangle\
.
\end{equation}
The Lagrangian density is:
\begin{equation}\label{4.5}
\mathcal{L}=\bar{\psi}(\imath\ls{\partial}-m(\sigma))\psi-
\frac{1}{2}{({\partial}_{\mu}\sigma)}^2-U(\sigma)\ ,
\end{equation}
with
\begin{equation}\label{4.6}
m(\sigma)=g\sigma\ .
\end{equation}
We choose $g=2$ for our calculations. The Euler-Lagrange equation
for the scalar field is:
\begin{equation}\label{4.7}
\frac{d^2\sigma}{d{\rho}^2}+\frac{1}{\rho}\frac{d\sigma}{d\rho}
=-\frac{\partial}{\partial\sigma}(P_{\psi}-U)\ .
\end{equation}
An appropriate form for the potential is:
\begin{equation}\label{4.8}
U=\frac{1}{4}\lambda{({\sigma}^2-{\sigma}_0^2)}^2.
\end{equation}
The proper generalization in order to include gravity in our
discussion is to replace the Fermi energy with a global chemical
potential, according to the equation \ref{4.14}, \cite{fr3}. In
this way the Fermi energy takes the proper generally covariant
form, being the zero component of the energy-momentum $4-$vector.
The generalization of the Lagrangian with the inclusion of
gravity is straightforward:
\begin{equation}\label{4.9}
\mathcal{L}/\sqrt{-g}=\frac{\imath}{2}(\bar{\psi}{\gamma}^{\mu}{\psi}_{;\mu}
-\bar{{\psi}_{;\mu}}{\gamma}^{\mu}\psi)
-m(\sigma)\bar{\psi}\psi+\frac{1}{2}{\sigma}_{;\mu}{\sigma}^{;\mu}-U(\sigma)\
.
\end{equation}
The proper rescalings for the case under discussion are made with
respect to a new mass scale $\mu$ equal to ${\sigma}_0^2$,
simplifying in this way the potential and extracting a ${\mu}^3$
factor from the Lagrangian. So the quantities in the Lagrangian
are rescaled in the following way:
\begin{equation}\label{4.10}
\tilde{\sigma}=\frac{\sigma}{{\sigma}_0}\ , \hspace{1em}
\widetilde{\psi}=\frac{\psi}{{\sigma}_0^2}\ , \hspace{1em}
\tilde{\lambda}=\frac{\lambda}{{\sigma}_0^2}\ , \hspace{1em}
\widetilde{m}=\frac{m}{{\sigma}_0^2}\ , \hspace{1em}
\tilde{\rho}=\rho{\sigma}^2_0\ , \hspace{1em}
\widetilde{\mathcal{L}}=\frac{\mathcal{L}}{{\sigma}^6_0}\ .
\end{equation}
We also make two new redefinitions:
\begin{equation}\label{4.11}
\tilde{r}=8\pi G{\sigma}^2_0\ , \hspace{1em}
\widetilde{\Lambda}=\frac{\Lambda}{8\pi G{\sigma}^6_0}\ .
\end{equation}

So, dropping the tildes we can regard that inside the soliton the
matter field derivative with respect to the radius is of order of
$O(G{\mu}^2)$ so the Euler-Lagrange equation for the scalar field
can be written:
\begin{equation}\label{4.12}
\frac{\partial}{\partial\sigma}(P_{\psi}-U)=0\ .
\end{equation}
The energy-momentum tensor is:
\begin{equation}\label{4.13}
T^{\mu\nu}=\frac{1}{2}\imath(\bar{\psi}{\gamma}^{(\mu}{\psi}^{;\nu)}-
{\bar{\psi}}^{;(\mu}{\gamma}^{\nu)}\psi)+
{\sigma}^{;\mu}{\sigma}^{;\nu}-g^{\mu\nu}[1/2{\sigma}_{;\alpha}{\sigma}^{;\alpha}-U(\sigma)]\
.
\end{equation}
We define a global chemical potential ${\omega}_{\psi}$ through
the equation:
\begin{equation}\label{4.14}
{\omega}_{\psi}^2={\varepsilon}_F^2B^{-1}(r)\ .
\end{equation}
This quantity retains a role similar to the frequency of the
other kinds of soliton stars. The particle number is:
\begin{equation}\label{4.15}
Q_{\psi}=\int\sqrt{-g}dx^1dx^2j^0=2\pi\int r
dr\sqrt{\frac{1}{A}}\langle{\psi}^{\dagger}\psi\rangle\ .
\end{equation}
The Einstein equations take the form:
\begin{equation}\label{4.16}
-\frac{1}{2r}\frac{dA}{dr}={\mathcal{E}}_{\psi}+U+\Lambda\ ,
\end{equation}
\begin{equation}\label{4.17}
-\frac{1}{2r}\frac{A}{B}\frac{dB}{dr}=P_{\psi}-U-\Lambda\ .
\end{equation}

Within the surface we repeat the discussion of eqs.
\ref{1.24}-\ref{1.27} and find from eq. \ref{4.12}:
\begin{equation}\label{4.18}
U=P_{\psi}\Rightarrow\frac{\lambda}{4}=\frac{1}{12\pi}{\omega}^3_{\psi}B^{3/2}_{\textrm{sur}}\
.
\end{equation}
This is the eigenvalue equation for the chemical potential. The
Einstein equations take the soluble form:
\begin{equation}\label{4.19}
-\frac{1}{2r}\frac{dA}{dr}=\frac{1}{6\pi}{\omega}^3_{\psi}B^{3/2}
+\frac{\lambda}{4}+\Lambda\ ,
\end{equation}
\begin{equation}\label{4.20}
-\frac{1}{2r}\frac{A}{B}\frac{dB}{dr}=\frac{1}{12\pi}{\omega}^3_{\psi}B^{3/2}
-\frac{\lambda}{4}-\Lambda\ ,
\end{equation}
The total charge is:
\begin{equation}\label{4.21}
Q=\frac{1}{2}\int drr{\omega}_{\psi}^2BA^{-1/2}\ .
\end{equation}
For our numerical calculations we use $\lambda=1/2$.

Outside the soliton we regard that $\sigma$ is everywhere unity
and $\psi$ zero, so both energy density and current are zero.

\begin{figure}
\centering
\includegraphics{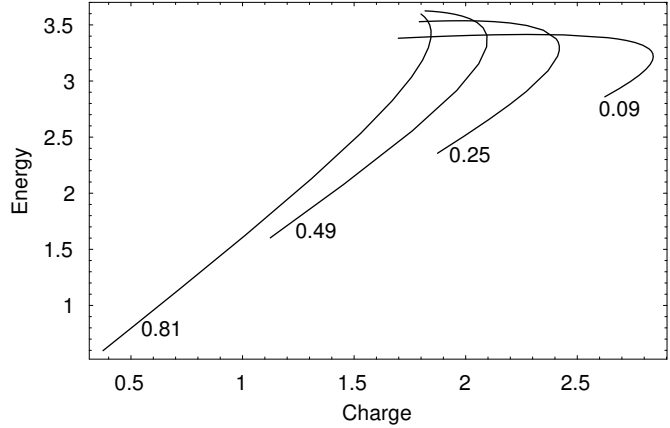}
\caption{Energy as a function of charge for four values of the
$A_{\textrm{sur}}$ or, equivalently, the chemical potential for a
fermionic q-star. We start from a small value of the cosmological
constant. Both energy and charge have small values. When
increasing the cosmological constant, in absolute values, both
energy and charge increase in order the soliton to generate
``positive" gravity as a counter effect to the ``negative" gravity
from the spacetime structure. But when $\Lambda$ takes very large
absolute values, a large soliton can not be stable and mass,
charge and radius decrease as one can see from figures
\ref{figure4.1}-\ref{figure4.4}.} \label{figure4.1}
\end{figure}

\begin{figure}
\centering
\includegraphics{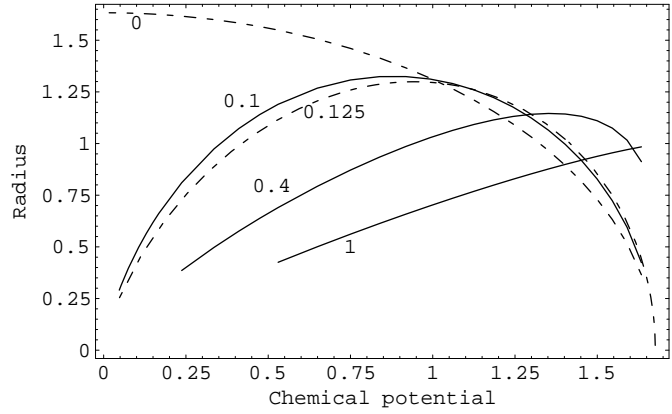}
\caption{The soliton radius as a function of the chemical
potential for a fermionic q-star for five values of the
cosmological constant. When ${\omega}_{\psi}$ is large the
gravity strength according to the eq. \ref{4.18} is small.}
\label{figure4.2}
\end{figure}

\begin{figure}
\centering
\includegraphics{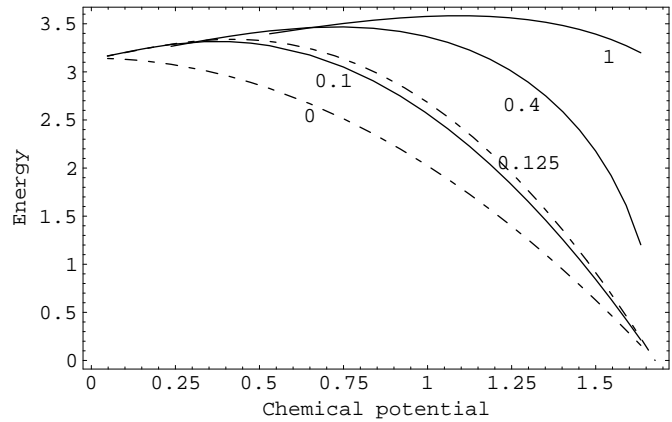}
\caption{The total soliton mass as a function of the chemical
potential for five values of the cosmological constant for a
fermionic q-star.} \label{figure4.3}
\end{figure}

\begin{figure}
\centering
\includegraphics{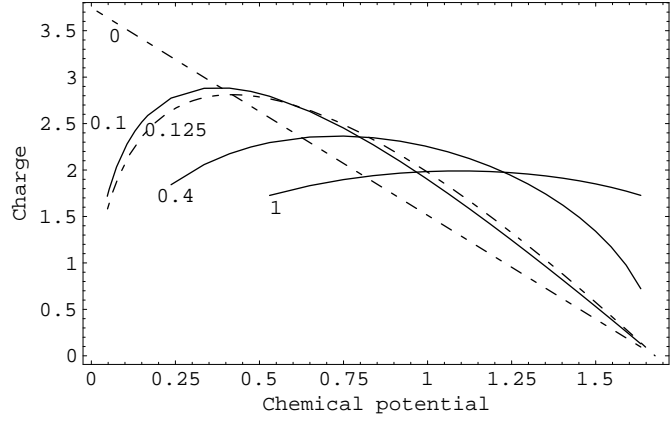}
\caption{The particle number as a function of the chemical
potential for five values of the cosmological constant for a
fermionic q-star.} \label{figure4.4}
\end{figure}

\begin{figure}
\centering
\includegraphics{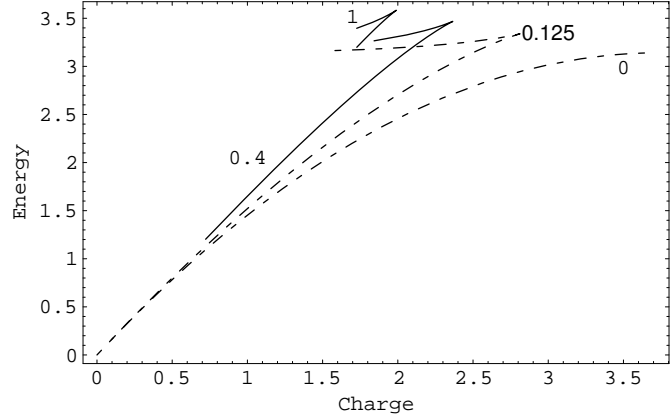}
\caption{The total mass of the field configuration as a function
of the charge for a fermionic q-star with constant cosmological
constant when varying the chemical potential. We start from large
values of this quantity (i.e: with $A_{\textrm{sur}}$ close to
unity, according to eq. \ref{4.18}) and, consequently, weak
gravity. Then both charge and energy have small values.
Decreasing ${\omega}_{\psi}$ both energy and charge increase up
to a certain value of the chemical potential, below which they
both decrease.} \label{figure4.5}
\end{figure}

Let us now take $\Lambda\rightarrow 0$. In that case the Einstein
equations admit the following analytical solution, also verified
numerically:
\begin{equation}\label{4.22}
B(r)=1/A_{\textrm{sur}}={\left(\frac{3\pi}{2}\right)}^{2/3}{\omega}_{\psi}^{-2}\
, \hspace{1em} A(r)=1-\frac{3}{8}r^2\ .
\end{equation}
We can easily find that the parameters of the soliton are:
\begin{equation}\label{4.23}
R={[(1-A_{\textrm{sur}})8/3]}^{1/2}= {\left[\left(1-
{\left(\frac{2}{3\pi}\right)}^{2/3}{\omega}_{\psi}^2\right)\frac{8}{3}\right]}^{1/2}\
,
\end{equation}
\begin{equation}\label{4.24}
E=(1-A_{\textrm{sur}})\pi=
\left[1-{\left(\frac{2}{3\pi}\right)}^{2/3}{\omega}_{\psi}^2\right]\pi\
,
\end{equation}
\begin{equation}\label{4.25}
Q={\left(\frac{3\pi}{2}\right)}^{2/3}\frac{4}{3}
(1-A_{\textrm{sur}}^{1/2})={\left(\frac{3\pi}{2}\right)}^{2/3}\frac{4}{3}
\left[1-{\left(\frac{2}{3\pi}\right)}^{1/3}{\omega}_{\psi}\right]\
,
\end{equation}
where $R$ is the soliton radius, $E$ the total mass and $Q$ the
soliton charge.

We will now examine a different case admitting analytical
solutions, namely:
\begin{equation}\label{4.26}
\Lambda=-\frac{\lambda}{4}
\end{equation}
In that case Einstein equations take the form
\begin{equation}\label{4.27}
-\frac{1}{2r}\frac{dA}{dr}=\frac{2}{12\pi}{\omega}^3_{\psi}B^{3/2}
\end{equation}
\begin{equation}\label{4.28}
-\frac{1}{2r}\frac{A}{B}\frac{dB}{dr}=\frac{1}{12\pi}{\omega}^3_{\psi}B^{3/2}
\end{equation}
We can find the soliton radius:
\begin{equation}\label{4.29}
R=4{(A_{\textrm{sur}}^{3/4}-A_{\textrm{sur}})}^{1/2}=
4\sqrt{-{\left(\frac{2}{3\pi}\right)}^{2/3}{\omega}_{\psi}^2
+{\left(\frac{2}{3\pi}\right)}^{1/2}{\omega}_{\psi}^{3/2}}\ .
\end{equation}
The metrics take the form:
\begin{eqnarray}\label{4.30}
A(r)=\frac{{(16A_{\textrm{sur}}-r^2+R^2)}^4}{65536A_{\textrm{sur}}^3}
\nonumber\\
=\frac{{(16A_{\textrm{sur}}^{3/4}-r^2)}^4}{65536A_{\textrm{sur}}^3}=
\frac{{\pi}^2{(3r^2-16\sqrt{6/\pi}{\omega}_{\psi}^{3/2})}^4}{2359296{\omega}_{\psi}^6}\
,
\end{eqnarray}
\begin{eqnarray}\label{4.31}
B(r)=\frac{1}{256}\frac{{(16A_{\textrm{sur}}-r^2+R^2)}^2}{A_{\textrm{sur}}^3}=
\nonumber\\
\frac{1}{256}\frac{{(16A_{\textrm{sur}}^{3/4}-r^2)}^2}{A_{\textrm{sur}}^3}=
\frac{{(3\pi
r^2-16\sqrt{6\pi}{\omega}_{\psi}^{3/2})}^2}{1024{\omega}_{\psi}^6}\
.
\end{eqnarray}
We can also find the energy and charge:
\begin{equation}\label{4.32}
E=(1+2A_{\textrm{sur}}^{3/4}-3A_{\textrm{sur}})\pi=
\pi-2^{2/3}{(3\pi)}^{1/3}{\omega}_{\psi}^2+2\sqrt{\frac{2\pi}{3}}{\omega}_{\psi}^{3/2}\
,
\end{equation}
\begin{equation}\label{4.33}
Q=2^{4/3}{(3\pi)}^{2/3}(A_{\textrm{sur}}^{1/4}-A_{\textrm{sur}}^{1/2})=
2\sqrt{6\pi}{\omega}_{\psi}^{1/2}-2^{5/3}{(3\pi)}^{1/3}{\omega}_{\psi}\
.
\end{equation}
For both the analytically soluble cases the chemical potential
${\omega}_{\psi}$ varies from the maximum value obtained by the
eq. \ref{4.18} (equal to $3\pi/2$ for these stars) when the
minimum value is obtained equating the soliton energy with the
energy of the free particles with the same charge. When
$B=A_{\textrm{sur}}^{-1}=\textrm{const.}$ this equation has no
solution and no decay to free particles is possible. In the latter
case the minimum value is ${\omega}_{\psi}=0.0481$.

We will now examine the $\Lambda+\lambda/4=0$ analytical solution
in more details because in the fermion-scalar soliton stars we
face the simpler relations when compared with those resulting
from the investigation of the solitons with two scalar fields.
With the help of figure \ref{figure4.5} we see that the $E=E(Q)$
function has two branches, a lower corresponding to large values
of $A_{\textrm{sur}}$, or, equivalently, ${\omega}_{\psi}$, and
the upper one corresponding to small values of the above
quantities. The boundary separating the two areas is determined
by the solution of the equations
$$\frac{dE}{d{\omega}_{\psi}}=0\ , \hspace{1em}
\frac{dQ}{d{\omega}_{\psi}}=0\ .$$ The solution to both of the
above equations is only one:
\begin{equation}\label{4.34}
{\omega}_{\psi}^{\textrm{cr}}=\frac{1}{4}{\left(\frac{3\pi}{2}\right)}^{1/3}\
, \hspace{1em} A_{\textrm{sur}}^{\textrm{cr}}=\frac{1}{16}\ .
\end{equation}
For ${\omega}_{\psi}>{\omega}_{\psi}^{\textrm{cr}}$ energy and
charge are monotone decreasing functions of the chemical
potential, but they are both monotone increasing for
${\omega}_{\psi}<{\omega}_{\psi}^{\textrm{cr}}$. The charge shows
a more rapid decrease.

Solving eq. \ref{4.33} in terms of $A_{\textrm{sur}}$ for
simplicity we find two solutions, called $A^{(1)}_{\textrm{sur}}$
and $A^{(2)}_{\textrm{sur}}$ corresponding to the upper and lower
branches respectively of the the $E=E(Q)$ function. These
solutions are:
\begin{eqnarray}\label{4.35}
A^{(1)}_{\textrm{sur}}=\frac{1}{72{\pi}^3}\left(2^{1/3}3^{2/3}Q^2{\pi}^{5/3}-
12\cdot2^{2/3}3^{1/3}Q{\pi}^{7/3}+36{\pi}^3-
6\sqrt{2}{\pi}^2\times \right. \nonumber\\ \left.
\sqrt{-2Q^3-12\cdot2^{2/3}3^{1/3}Q{\pi}^{4/3}
+18{\pi}^2+5\cdot2^{1/3}Q^2{(3\pi)}^{2/3}}\right)
\end{eqnarray}
\begin{eqnarray}\label{4.36}
A^{(2)}_{\textrm{sur}}=\frac{1}{72{\pi}^3}\left(2^{1/3}3^{2/3}Q^2{\pi}^{5/3}+
12\cdot2^{2/3}3^{1/3}Q{\pi}^{7/3}+36{\pi}^3-
6\sqrt{2}{\pi}^2\times \right. \nonumber\\ \left.
\sqrt{-2Q^3-12\cdot2^{2/3}3^{1/3}Q{\pi}^{4/3}
+18{\pi}^2+5\cdot2^{1/3}Q^2{(3\pi)}^{2/3}}\right)
\end{eqnarray}
The metric of the eq. \ref{4.35} is the small one (less than
$1/16$) when the the metric of eq. \ref{4.36} is the large one.
Substituting in eq. \ref{4.32} we find the energies corresponding
to the two branches:
\begin{eqnarray}\label{4.37}
E^{(1)}=-\frac{\pi}{2}+Q{\left(\frac{2\pi}{2}\right)}^{1/3}-
\frac{Q^2}{4\cdot2^{3/2}{(3\pi)}^{1/3}}+ \nonumber\\ \frac{1}{4}
\sqrt{-4Q^3-24\cdot2^{2/3}3^{1/3}Q{\pi}^{4/3}+36{\pi}62+10\cdot2^{1/3}Q^2{(3\pi)}^{2/3}}+
\nonumber\\
\frac{1}{6\cdot2^{1/4}3^{1/2}{\pi}^{5/4}}
\left(2^{1/3}3^{2/3}Q^2{\pi}^{5/3}-12\cdot2^{2/3}3^{1/3}Q{\pi}^{7/3}+
  6{\pi}^2\times \right. \nonumber\\ \left. (6\pi-
\sqrt{-4Q^3-24\cdot2^{2/3}3^{1/3}Q{\pi}^{4/3}+36{\pi}^2+10\cdot2^{1/3}Q^2{(3\pi)}^{2/3}})
\right)^{3/4},
\end{eqnarray}
the large value of the energy corresponding to the upper branch
and:
\begin{eqnarray}\label{4.38}
E^{(2)}=-\frac{\pi}{2}+Q{\left(\frac{2\pi}{2}\right)}^{1/3}-
\frac{Q^2}{4\cdot2^{3/2}{(3\pi)}^{1/3}}+ \nonumber\\ \frac{1}{4}
\sqrt{-4Q^3-24\cdot2^{2/3}3^{1/3}Q{\pi}^{4/3}+36{\pi}62+10\cdot2^{1/3}Q^2{(3\pi)}^{2/3}}+
\nonumber\\
\frac{1}{6\cdot2^{1/4}3^{1/2}{\pi}^{5/4}}
\left(2^{1/3}3^{2/3}Q^2{\pi}^{5/3}-12\cdot2^{2/3}3^{1/3}Q{\pi}^{7/3}+
  6{\pi}^2\times  \right. \nonumber\\ \left.  (6\pi+
\sqrt{-4Q^3-24\cdot2^{2/3}3^{1/3}Q{\pi}^{4/3}+36{\pi}^2+10\cdot2^{1/3}Q^2{(3\pi)}^{2/3}})
\right)^{3/4},
\end{eqnarray}
the small one.

\section{Conclusions}

Our discussion can be summarized as follows:

1.  There are stable, gravitating field configurations, in $2+1$
dimensions, consisted of a $N-$carrying field, fermionic, or
scalar (abelian or non-abelian). An always positive potential,
generated either by the the $N-$carrying field, or by an
additional real scalar field, is necessary in order to stabilize
the soliton. The potential as a function of the field $\phi$
increases as ${\phi}^2$ for small values of the field, then
slower than ${\phi}^2$ providing in this way a local minimum for
some value of the field differing from zero and for large values
of the field as ${\phi}^4$, or ${\phi}^6$ usually, in order to
ensure the positivity of the potential for large values of the
field. The soliton solution lies near this minimum and the total
energy of the star is smaller than $m^2{\phi}^2$, i.e. the
potential energy of the free particles in any field configuration
under consideration . There is always a certain frequency,
minimizing the total energy of the configuration, obtained by the
equality of the $W-$energy to the $U-$energy. This equality holds
true in the absence of gravity and is generalized in a generally
covariant way.

2. When $\Lambda\neq0$ and $A_{\textrm{sur}}\rightarrow0$ the
particle number decreases rapidly but the total energy decreases
slowly, as one can see from figures \ref{figure1.1},
\ref{figure2.1}, \ref{figure3.1} and \ref{figure4.5}. So, the
$A_{\textrm{sur}}=0$ case (formation of horizon) is impossible,
because below a certain value of $A_{\textrm{sur}}$ fission into
free particles is energetically favorable. This result is
supported by both numerical and analytical solutions. Decay into
free particles is energetically forbidden for zero cosmological
constant, as can be proved by the analytical solutions for every
special kind of q-star. For cosmological constant differing from
zero, the energy and the charge of the soliton increase when the
gravity becomes stronger (i.e.: when $A_{\textrm{sur}}$
decreases) but below a certain value of the frequency energy
decreases slowly and charge more rapidly. The result is that for
a certain charge there are usually two different values of the
energy. So the soliton with the larger energy can emit the energy
excess and goes to the lower-energy state. Due to the above
behavior of the $E=E(Q)$ function, we found analytically and
numerically that below a certain value of the frequency the
soliton decay to free particles is energetically favorable, so
when the cosmological constant differs from zero there is a
certain range of the soliton parameters (the independent
parameter, frequency, and the derivative ones, radius, energy and
charge) which verify the soliton stability. This range of the
phase space parameters is fully depicted in our figures.

3.  When the cosmological constant is zero or small with respect
to the energy densities (namely $U$ and $W$), the soliton
parameters, radius, mass and particle number, increase when the
frequency decreases, i.e. when the gravity becomes stronger. When
the cosmological constant becomes more important than the energy
densities, the soliton becomes smaller and the mass and particle
number show a corresponding decrease.

\vspace{1em}

\textbf{Acknowledgments}

\vspace{1em}

I wish to thank N.D. Tracas and E. Papantonopoulos for helpful
discussions.

\end{document}